%% file: main_arxiv.tex
\documentclass[amssymb,prb,twocolumn,superscriptaddress,floats,showkeys,showpacs]{revtex4-2}
\usepackage[T1]{fontenc}
\usepackage{bm}
\usepackage{graphicx}
\usepackage{amssymb}
\usepackage{leftidx}
\usepackage{upgreek}
\usepackage{siunitx}
\usepackage{amsfonts}
\usepackage{amsmath} 
\usepackage{hyperref} 
\usepackage{multirow}
\usepackage{makecell}
\usepackage{xcolor}
\usepackage{xspace}
\hypersetup{colorlinks,citecolor=blue, filecolor=blue ,linkcolor=blue , urlcolor=blue, pdftex}

\usepackage{tabularx}

\begin{document}

\newcommand{\WSe}{$\text{WSe}_{2}$\xspace}
\newcommand{\CGT}{$\text{Cr}_{2}\text{Ge}_{2}\text{Te}_{6}$\xspace}

\title{Electrically modulated light-emitting diodes driven by resonant and antiresonant tunneling between Cr$_2$Ge$_2$Te$_6$ electrodes}

% {author & affiliation}

\def \FUW{University of Warsaw, Faculty of Physics, 02-093 Warsaw, Poland}

\def \sgg{Department of Materials Science and Engineering, National University of Singapore, 117575, Singapore} 
\def \sg{Institute for Functional Intelligent Materials, National University of Singapore, 117544, Singapore}

\def \Watanabe{Research Center for Electronic and Optical Materials, National Institute for Materials Science, 1-1 Namiki, Tsukuba 305-0044, Japan}
\def \Taniguchi{Research Center for Materials Nanoarchitectonics, National Institute for Materials Science, 1-1 Namiki, Tsukuba 305-0044, Japan}

\def \Leipzig{ Institute of Inorganic Chemistry, Faculty of Chemistry and Mineralogy, Leipzig University, 04103 Leipzig, Germany}

\author{Natalia Zawadzka}
\email{Natalia.Zawadzka@fuw.edu.pl}
\affiliation{\FUW}
\author{Kristina Vaklinova}
\affiliation{\sg}
\author{Tomasz Woźniak}
\affiliation{\FUW}
\author{Mihai I. Sturza}
\affiliation{\Leipzig}
\author{Holger~Kohlmann}
\affiliation{\Leipzig}
\author{Kenji~Watanabe}
\affiliation{\Watanabe}
\author{Takashi Taniguchi}
\affiliation{\Taniguchi}
\author{Adam Babiński}
\affiliation{\FUW}
\author{Maciej Koperski}
\affiliation{\sg}
\affiliation{\sgg}
\email{msemaci@nus.edu.sg}
\author{Maciej R. Molas}
\email{Maciej.Molas@fuw.edu.pl}
\affiliation{\FUW}

\begin{abstract} 

Exploring the electron tunneling mechanisms in diverse materials systems constitutes a versatile strategy for tailoring the properties of optoelectronic devices. In this domain, bipolar vertical tunneling junctions composed of van der Waals materials with vastly different electronic band structures enable simultaneous injection of electrons and holes into an optically active material, providing a universal blueprint for light-emitting diodes (LEDs). Efficient modulation of the injection efficiency has previously been demonstrated by creating resonant states within the energy barrier formed by the luminescent material. Here, we present an alternative approach towards resonant tunneling conditions by fabricating tunneling junctions composed entirely from gapped materials: Cr$_2$Ge$_2$Te$_6$ as electrodes, hBN as a tunneling barrier, and monolayer WSe$_2$ as a luminescent medium. The characterization of such LEDs revealed a nonmonotonous evolution of the electroluminescence intensity with the tunneling bias. The dominant role driving the characteristics of the electron tunneling was associated with the relative alignment of the density of states in Cr$_2$Ge$_2$Te$_6$ electrodes. The unique device architecture introduced here presents a universal pathway towards LEDs operating at room temperature with electrically modulated emission intensity.

\end{abstract}

\maketitle

%%%%%%%%%%%%%%%%%%%%%%%%%%%%%%%%%%%%%%%%%%%%%%%%%%%%%%%%%%%%%%%%5
\section{Introduction \label{sec:Intro}}

Vertical tunneling junctions assembled as van der Waals (vdW) stacks constitute a universal template for the rational design of bipolar light-emitting diodes (LEDs). Typically, three types of materials are utilized to inject electrons and holes into the luminescent states: 1) (semi-)metallic electrodes acting as a source of charge carriers, 2) large bandgap insulators creating tunneling barriers, and 3) semiconductor systems fulfilling the role of an optically active medium. 
The advantage of the vdW fabrication technology leverages vertical stacking of atomically thin layers of materials beyond the limitations of matching crystal structures or lattice constants, enabling broad tunability of the optoelectronic properties of vdW LEDs. 
The excitonic energy~\cite{WSe2_excitons, MoSe2_excitons, molasNanoscale} governs the emission wavelength in semiconducting or insulating materials ($e.g.$, in various representatives of transition metal dichalcogenides (TMDs)~\cite{TMD_excitons_review, Mak2010, Arora2015W, Arora2015Mo, Lezama2015, Wang_2018, molasNanoscale}). Further control arises from confinement effects strongly dependent on the exact number of layers ($e.g.$, in InSe quantum wells~\cite{InSe_LED, InSe_confinement}). 
The incorporation of sparse defect centers~\cite{WSe2_SPE, hBN_SPE, hBN_defect_STM, hBN_quantum_emitters, WS2_Nb_doped} enables the electrical excitation of quantum emitters~\cite{Grzeszczyk2024, WSe2_quantum_LED, LED_quantum_perspective, 2D_SPE_review}, while the introduction of ferromagnetic electrodes unlocks spin injection schemes~\cite{Ye2016, Li2022, Zhang2022}.

The efficiency of the electroluminescent processes in vdW LEDs strongly depends on the charge carrier dynamics, when the tunneling events compete with radiative and non-radiative recombination occurring at the states participating in the optical transitions. This tunneling dynamics has been demonstrated to be affected by creating resonant conditions within optically active materials, $e.g.$, by utilizing bound states within a quantum well, or by introducing discrete defect levels isolated from the band edges~\cite{hBN_carbon_vacancy, hBN_defects_ab_inito, hBN_radiative_defects, hBN_dielectric, Grzeszczyk2024}.

Here, we demonstrate an alternative route towards the realization of resonant tunneling by introducing a gapped Cr$_2$Ge$_2$Te$_6$ semiconductor as an electrode material, in conjunction with commonly used hBN as a tunneling barrier and monolayer (ML) WSe$_2$ as the luminescent medium. 
The intrinsic doping in Cr$_2$Ge$_2$Te$_6$ provides free carriers to participate in the tunneling processes at room temperature, enabling electrical excitation of excitons characterized by a large binding energy in WSe$_2$ MLs. 
Unlike vdW LEDs with graphene electrodes, we found that the intensity of electroluminescence (EL) evolves nonmonotonously with the increasing tunneling bias. 
This behavior constitutes a hallmark of the relative alignment of resonances in the density of states (DOS) in Cr$_2$Ge$_2$Te$_6$ arising from multiple valence and conduction sub-bands, which strongly modulates the efficiency of the tunneling process. 
The resonant conditions, when the maxima in Cr$_2$Ge$_2$Te$_6$ DOS occur in the final state of the tunneling process, favor fast nonradiative tunneling pathways through the vertical junction, which significantly decreases the EL intensity. 
The antiresonant conditions, which involve Cr$_2$Ge$_2$Te$_6$ DOS minima in the final state of the tunneling process, redirect the tunneling electrons towards radiative pathways. 

The vdW tunneling LED architecture, based entirely on gapped material, enables novel functionalities of modulating EL processes by leveraging unique material properties. Multiple subbands emerging in Cr$_2$Ge$_2$Te$_6$ layers participate in the tunneling events, which lay the foundation for a novel class of ultrathin devices with electrically excited and modulated emission intensity operating at room temperature.

%%%%%%%%%%%%%%%%%%%%%%%%%%%%%%%%%%%%%%%%%%%%%%%%%%%%%%%%%%%%%%
\section{Experimental results\label{sec:Results}}

\begin{figure}[]
    \centering
    \includegraphics[width=1\linewidth]{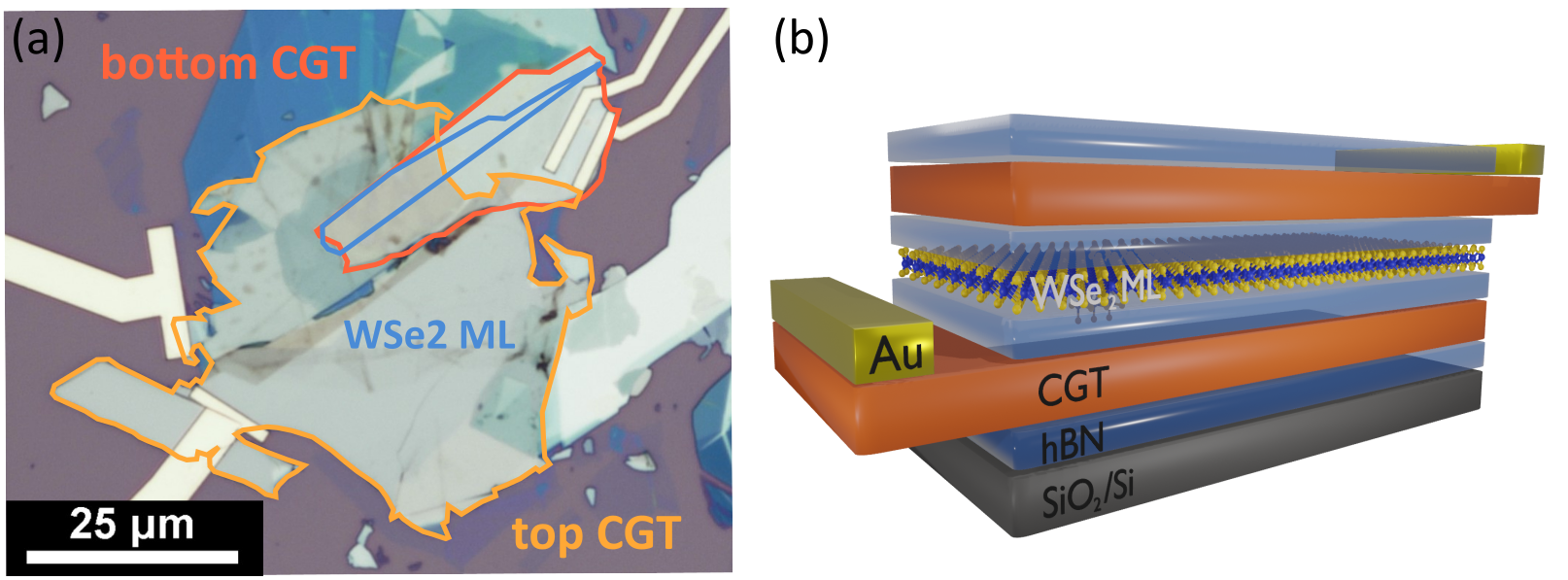}
    \caption
    {\label{sheme}
    The structure of the light-emitting diode (LED) based on van der Waals (vdW) tunneling junction. (a) Microscope optical image of the active part of the Cr$_2$Ge$_2$Te$_6$/hBN/WSe$_2$/hBN/Cr$_2$Ge$_2$Te$_6$ vdW LED. The \WSe monolayer and Cr$_2$Ge$_2$Te$_6$ contacts are highlighted by the outlines. (b) Schematic drawing of the vdW stack used in the LED architecture encapsulated with hBN films and deposited onto Si/SiO$_2$ wafer.}
\end{figure}

We fabricated a vdW LED with Cr$_2$Ge$_2$Te$_6$/hBN/WSe$_2$/hBN/Cr$_2$Ge$_2$Te$_6$ architecture encapsulated with hBN films on a Si/SiO$_2$ wafer. The optical image of the device and the structure of the vdW stack are presented in Fig.~\ref{sheme}. 
The CGT contacts have different thicknesses of about 20 nm and 75 nm for the bottom and top contacts, respectively. 
The hBN thicknesses are in the range of a few layers, allowing efficient carrier tunneling~\cite{Withers2015, Withers2015WSe2, Binder2017, WALCZYK2025}.
The crystal growth methodology and the fabrication recipe of the vdW LED are presented in the Methods section.

\begin{figure}[t]
    \centering
    \includegraphics[width=0.8\linewidth]{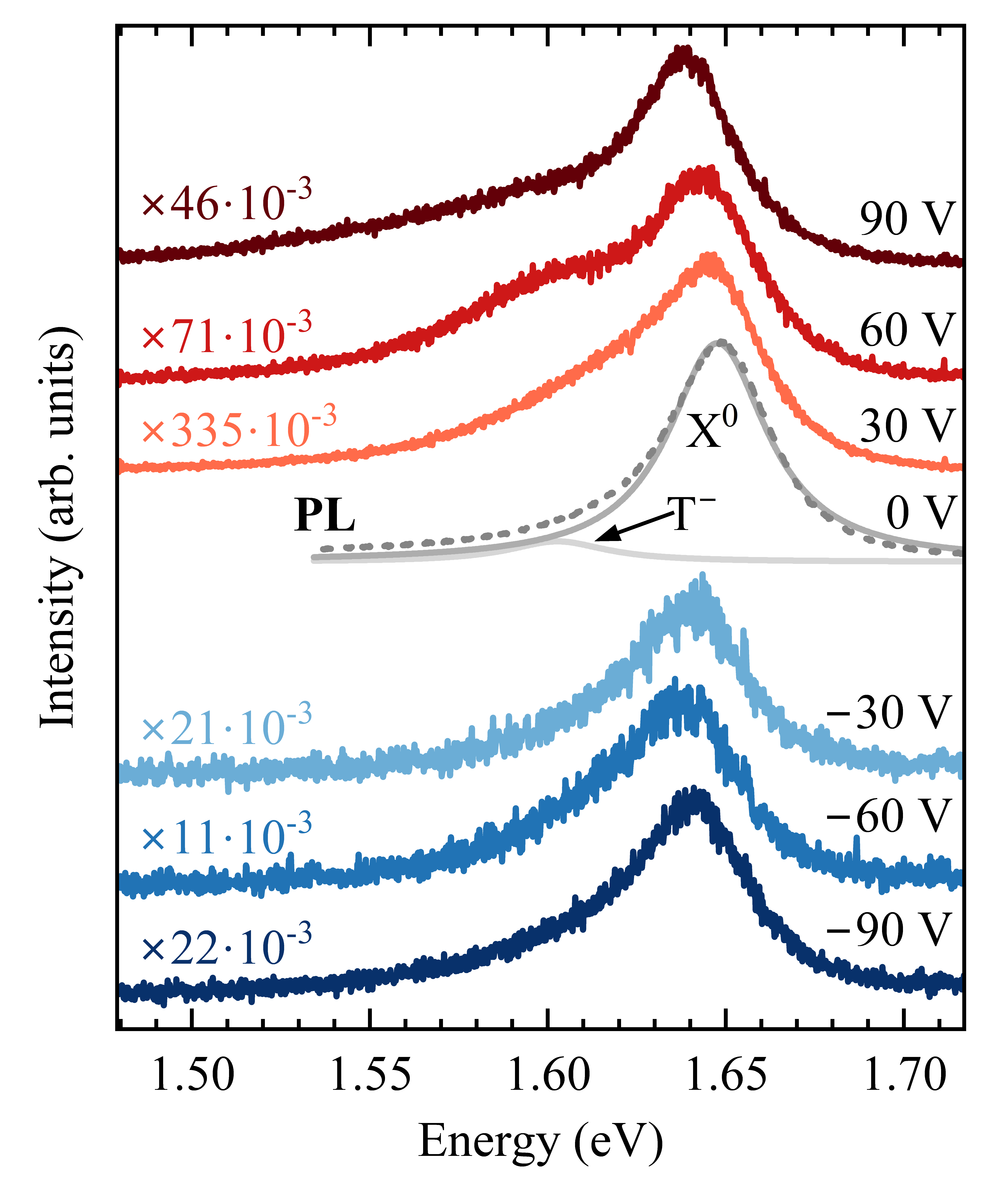}
    \caption
    {\label{Fig_2}  
    The room temperature ($\mathrm{T~=~300~K}$) photoluminescence (PL) and electroluminescence (EL) spectra of the Cr$_2$Ge$_2$Te$_6$/hBN/WSe$_2$/hBN/Cr$_2$Ge$_2$Te$_6$ light-emitting diode measured in a microscopic configuration. 
    The PL spectrum was obtained at zero bias voltage and the EL spectra were detected at tunnelling bias voltages $V_T = \pm 30,~60,~90~V$. 
    The EL spectra are multiplied by scaling factors shown in the figure for ease of comparison. The spectra are vertically shifted for clarity. Two Lorentz functions are fitted to the PL spectrum for the identified excitonic complexes, $i.e.$, X$^0$ and T$^-$.}  
\end{figure}

The optoelectronic properties of the vdW LED device were examined at room temperature ($\mathrm{T~=~300~K}$). 
The photoluminescence (PL) and EL spectra are presented comparatively in Fig.~\ref{Fig_2}. The PL response is dominated by a broad resonance centered at 1.649~eV with a visible tail at lower energies. Two Lorentz functions were fitted to determine the spectral position of both features. 
The dominant higher-energy emission line is attributed to a neutral free exciton (X$^0$) in the vicinity of the band gap in the K$^+$/K$^-$ valleys of the Brillouin zone (BZ), while a significantly less intense line at the lower-energy side is attributed to a negative trion (T$^-$)~\cite{Jones2013, Wang2014, Courtade2017, Withers2015WSe2, Li2019, Zinkiewicz2022}. 
The negative charge state of the T$^-$ trion was identified based on the PL spectra measured at a low temperature ($\mathrm{T~=~5~K}$), as demonstrated in Fig.~S1 in the Supplementary Information (SI).
This indicates that the investigated WSe$_2$ MLs was n-doped, which is commonly observed for WSe$_2$ intrinsic dopants.~\cite{Jones2013, Courtade2017, Withers2015, Li2019, Zinkiewicz2022, WALCZYK2025}. 

\begin{figure*}[!t]
    \centering
    \includegraphics[width=1\linewidth]{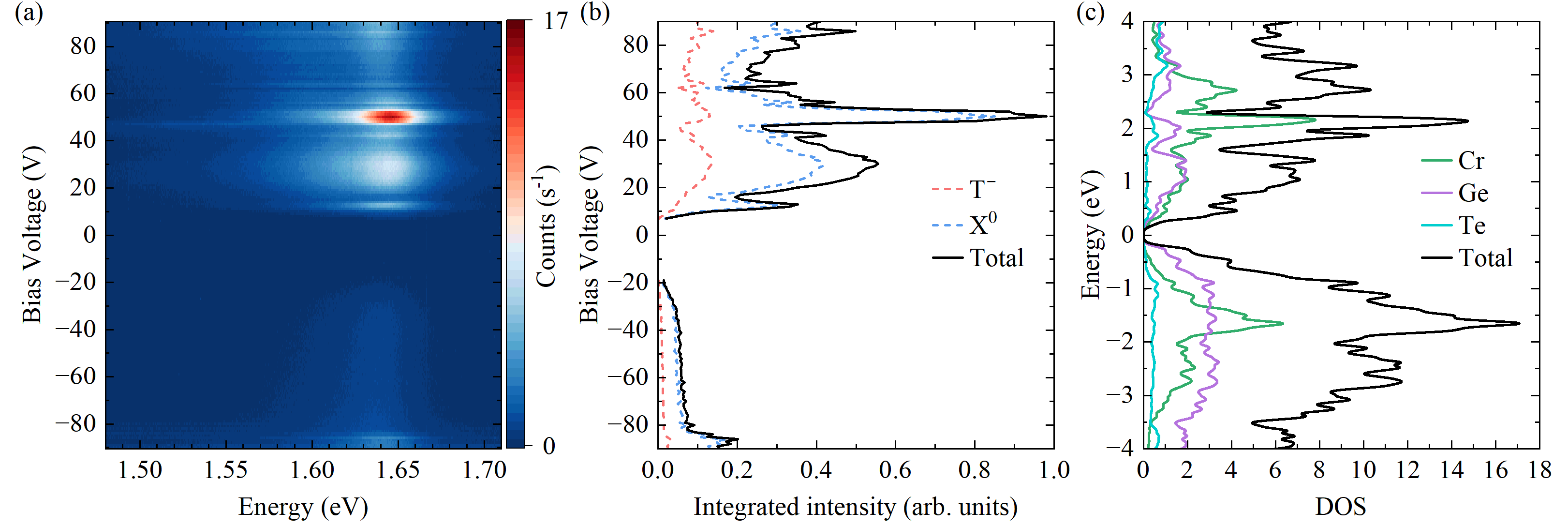}
    \caption
    {\label{Fig_3}
    (a) False-color map presenting the electroluminescence spectrum as a function of the applied voltage in the range of -90 V to 90 V. (b) Voltage-dependent integrated intensity of the EL signal with the corresponding X$^0$ and T$^-$ intensities in the same bias range as in panel (a). (c) Calculated density of states of Cr$_2$Ge$_2$Te$_6$ resolved by the atomic contributions from Cr, Ge, and Te. The energy scale in panel (c) is chosen to facilitate the comparison with the results shown in panel (b).
    }
\end{figure*}

The EL spectra are qualitatively similar to the PL spectra, comprising both X$^0$ and T$^-$ excitons. 
However, there are three main differences between the PL and EL spectra:
(1) The PL intensity measured at zero bias is more than an order of magnitude stronger than the EL intensity.
PL measured on a 1~$\mathrm{\mu m^2}$ spot using 2.21~eV excitation with a power of 20~$\mu$W corresponds to an excitation rate of $5.7 \times 10^{13}$ photons per second. 
For EL, the measured current of 32.2~$\mu$A at bias voltage 30~V corresponds to excitation of $2\times10^{14}$ electrons per second for the entire LED structure (the estimated area of about 71~$\mathrm{\mu m^2}$).
Since we collect the signal from the 1~$\mathrm{\mu m^2}$ area, the excitation rate of the electrons is reduced to $2.8\times10^{12}$ electrons per second.
The integrated EL signal at 30~V is 0.47 counts per second, while for PL it is 28 counts per second. 
Finally, we obtain a photon-to-photon conversion of about $2.1 \times 10^{12}$ and an electron-to-photon conversion of around $6.1 \times 10^{12}$. 
This implies that the electron-to-photon conversion was about 2.9 times smaller than the photon-to-photon conversion. 
(2) Although the PL intensity scales with the laser excitation power (see Fig.~S2 in the SI), the dependence of the EL intensity on the tunneling bias is nonmonotonic.
This contrasts with LED devices employing graphene layers as semimetallic electrodes, which exhibit stronger EL intensity with increasing tunneling bias and current~\cite{Ross2014, Withers2015, Withers2015WSe2, Lien2018, WALCZYK2025}.
(3) The relative intensity of the charged and neutral exciton emissions is bias-dependent and exhibits nonmonotonic behavior.
The X$^0$/T$^-$ ratio varies between 2 and 7.4 (see Fig.~S3 in the SI).
Excitation-power-dependent PL measurements, presented in the SI, reveal a monotonic change of the X$^0$/T$^-$ ratio in the range from 2.8 to 3.3.
In the bias voltage range from 60 to 90~V, the T$^-$ intensity is significantly enhanced and can be clearly resolved in the EL spectra (with an X$^0$/T$^-$ ratio of about 2.5), while at other voltages the X$^0$ emission is much stronger (with the ratio reaching values as high as 7).
This observation suggests an imbalance in the tunneling of electrons and holes in the investigated LED, such that the charge state of the WSe$_2$ monolayer becomes dependent on the tunneling bias.

\begin{figure*}[]
    \centering
    \includegraphics[width=1\linewidth]{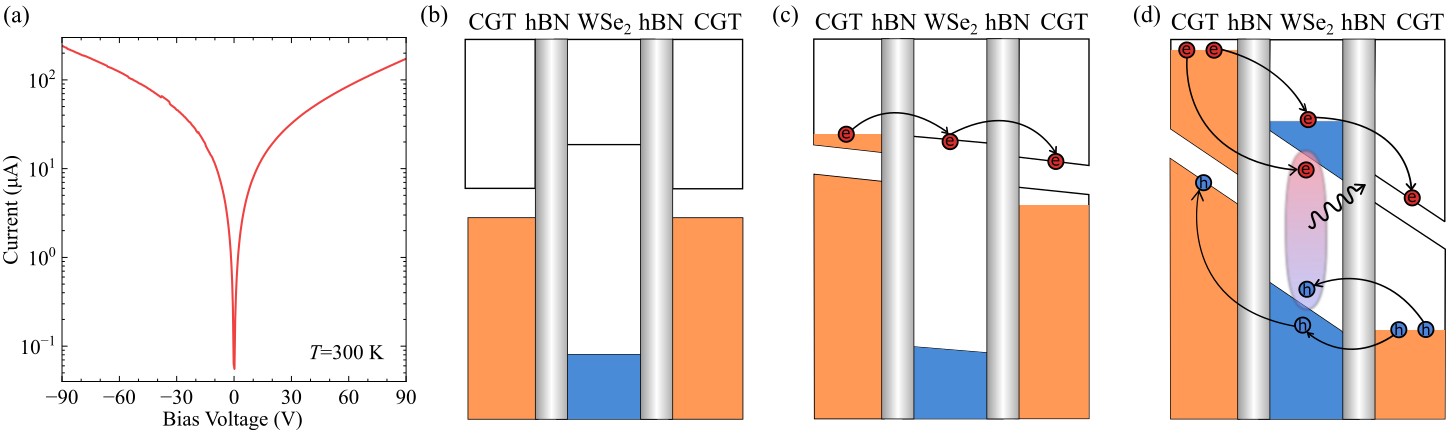}
    \caption
    {\label{fig_4}
    (a) Tunneling current–voltage characteristics of the studied light-emitting diode device, shown on a logarithmic scale.
    (b) Schematic illustration of the type-II band alignment in the Cr$_2$Ge$_2$Te$_6$/hBN/WSe$_2$/hBN/Cr$_2$Ge$_2$Te$_6$ heterostructure. The occupied states in Cr$_2$Ge$_2$Te$_6$ are shown in orange, while the occupied states in WSe$_2$ are shown in blue. The hBN barriers are represented by gray rectangles.
    (c) Applied intermediate bias below the electroluminescence threshold (from –19 V to 7 V). In this regime, electrons can tunnel through the hBN barrier into WSe$_2$, while holes cannot tunnel due to the absence of available empty states.
    (d) At higher bias voltages, both electrons and holes can tunnel into WSe$_2$ and form excitons, which recombine radiatively, resulting in electroluminescence. Alternatively, they may tunnel further into the opposite Cr$_2$Ge$_2$Te$_6$ contact without radiative recombination.}   
\end{figure*}

To investigate the aforementioned nonlinearity effects, we present the false color map of the EL spectra as a function of the tunneling bias in Fig.~\ref{Fig_3}(a). 
Multiple features are apparent in the EL evolution with bias, including narrow maxima observed at 13~V and 42~V and a broad resonance at around 30~V. Such behavior of the EL intensity points towards additional tunneling mechanisms beyond the increasing number of electrons and the activation of novel tunneling pathways mediated via bound states in the luminescent material, which are observed in vdW LED devices utilizing graphene as an electrode~\cite{Ross2014, Withers2015, Withers2015WSe2, Lien2018, WALCZYK2025}. One of the major differences emerging from semiconducting Cr$_2$Ge$_2$Te$_6$ electrodes is related to the density of states (DOS) around the conduction band edges. In graphene-based devices, usually a singular valence and conduction bands characterized by a linear dispersion participate in the tunneling events~\cite{Withers2015, Withers2015WSe2, Binder2017, Lien2018, Wang2018, Mueller2018, WALCZYK2025}, while multiple conduction and valence sub-bands determine the DOS in Cr$_2$Ge$_2$Te$_6$ around the band edges~\cite{WangCGT2018,Li2018, Kang2019, Suzuki2019}. In vdW LED devices utilizing Cr$_2$Ge$_2$Te$_6$, the DOS in the electrodes will become misaligned due to the presence of an electric field, which shifts their relative band edges. As the probability of the tunneling process scales with the number of available final states in a tunneling event due to the fermionic nature of electrons, the relative displacements of the DOS in Cr$_2$Ge$_2$Te$_6$ electrodes will create resonant (antiresonant) tunneling conditions, when the DOS maxima in the electrode with occupied states are aligned with maxima (minima) of the electrode with empty states. As the tunneling probability determines the average tunneling timescales, the lifetime of electrons and holes injected into the luminescent states of the WSe$_2$ monolayer will become bias dependent. In such conditions, the intensity of the EL signals will be modulated by the competition between non-radiative direct tunneling across the junctions and the radiative tunneling when electrons and holes live long enough in the WSe$_2$ monolayer to recombine. Additional mechanisms related to the modulation of the EL intensity can be related to non-radiative recombination processes ($e.g.$, via Auger mechanisms)~\cite{Binder2019}. The injection of charge carriers to higher-energy valence and conduction sub-bands can, in principle, impact the efficiency of Auger processes, making the non-radiative mechanisms intrinsic to WSe$_2$ monolayer bias-dependent.

In contrast, the EL evolution at negative voltages experiences a simpler behavior, dominated mainly by a monotonic increase in the EL intensity up to almost -90~V. 
Fig.~\ref{Fig_3}(b) presents the bias dependence of the integrated total EL intensity in conjunction with the fit intensity of the neutral exciton and the negative trion. Firstly, we can see a large difference in the bias voltages at which the EL signal emerges at the positive and negative sides. EL emission appears at around 7~V for positive voltages, whereas much higher voltages of about $-19$~V must be applied to raise EL signals for negative voltages. This behavior indicates an asymmetry arising within our vdW LED, which differentiates between the two Cr$_2$Ge$_2$Te$_6$ electrodes. This may be attributed to the different thicknesses of the electrodes, which affect the relative alignment of the DOS at positive and negative biases, and the proximity to the substrate~\cite{WALCZYK2025, Grzeszczyk2024}. In particular, the doping level of the Cr$_2$Ge$_2$Te$_6$ layers can be different in the top and bottom electrodes, due to varied proximity to the Si/SiO$_2$ substrate, which determines the efficiency of the charge transfer between the Cr$_2$Ge$_2$Te$_6$ layers and the Si substrate~\cite{Dolui2013,Grzeszczyk2021, Jadczak_2021}.

Let us now focus on the analysis of the intensities of X$^0$ and T$^-$ as a function of the applied bias voltage, which are demonstrated in Fig.~\ref{Fig_3}(b). It is seen that the X$^0$ intensity is much larger than the T$^-$ in the entire voltage range from -90~V to 90~V. The qualitative dependence of the X$^0$ and T$^-$ intensity on the bias is similar. However, the exact ratio between the intensity of the two resonances changes with bias. For example, the maximum of the X$^0$ intensity at about 50~V is not equally pronounced for the T$^-$ resonance. This indicates that the rate of electron and hole injection is different, effectively changing the charge state of the WSe$_2$ monolayer, which remains negatively charged in the investigated range of voltages.

To verify that the DOS plays a critical role in the optoelectric properties of a vdW LED with Cr$_2$Ge$_2$Te$_6$ electrodes, we compared the bias-dependent EL intensity with the atom-resolved DOS calculated at the density functional level, which is presented in Fig.~\ref{Fig_3}(c). The technical details of the calculations are discussed in the Methods section. Note that the energy scale in Fig.~\ref{Fig_3}(c) was adjusted from -4~eV to 4~eV to facilitate its comparison with the voltage evolution of the EL signal shown in panel (b) of the figure. We found that the integrated EL intensity on the positive bias side closely follows the calculated DOS of the Cr$_2$Ge$_2$Te$_6$. This indicates that the resonant and anti-resonant tunneling conditions are mostly responsible for the evolution of the EL intensity with bias~\cite{Chang1974, VanHoof1992, Bertram1994}. On the other hand, on the negative bias side, the EL intensity remains mostly unrelated to the DOS. This observation, in conjunction with the significant quenching of the EL intensity when compared to the positive bias side, indicates a dominant role of the nonradiative processes intrinsic to the WSe$_2$ monolayer.

The current-voltage (\textit{IV}) characteristics, demonstrated in Fig.~\ref{fig_4}(a), further corroborate our interpretation of the role of the resonant tunneling and nonradiative recombination in the operation of LED devices. The tunneling current as a function of the applied voltage undergoes a gradual increase with measurable currents of about 50~nA in the vicinity of the zero bias. This indicated that the number of electrons and holes tunneling through the junction is monotonously increasing, therefore, the modulation of the EL intensity needs to be related to the charge carrier dynamics and competition between various tunneling pathways.

The low onset of the tunneling current contrasts with the emergence of the EL signal at a much higher voltage. This is related to the alignment between the band edges of Cr$_2$Ge$_2$Te$_6$ and WSe$_2$ monolayers, as well as a large DOS in Cr$_2$Ge$_2$Te$_6$, which implies smaller changes to the Fermi level with bias when compared to, $e.g.$, graphene. In addition, the asymmetry in the \textit{IV} curve can also be observed. The tunneling current reaches 273 $\mu$A at -90~V, while it is almost 40$\%$ smaller at 90~V with a value of 173~$\mu$A. The overall behavior of the optoelectronic properties of vdW LEDs with Cr$_2$Ge$_2$Te$_6$ electrodes can be summarized as a result of the type II band alignment of the Cr$_2$Ge$_2$Te$_6$/hBN/\WSe/hBN/Cr$_2$Ge$_2$Te$_6$ heterostructure, as shown in Fig.~\ref{fig_4}(b).  First, we determine the relative alignment of the conduction band (CB) minima and the valence band (VB) maxima between the Cr$_2$Ge$_2$Te$_6$ electrodes and \WSe ML. The reported values for the CB minimum of ML \WSe relative to the vacuum level range from 3.53 to 3.56~eV, while the position of the VB maximum lies in the range of 4.86-5.16~eV~\cite{Kang2013, Gong2013}. For Cr$_2$Ge$_2$Te$_6$, the alignment of the band with respect to the vacuum level has been reported only for ML~\cite{Ko2022}. However, the band gap and the work function of Cr$_2$Ge$_2$Te$_6$ are highly dependent on the number of layers~\cite{Yimei2018, Rahman2021, Ko2022, Zhang2022}. Due to the thicknesses of the Cr$_2$Ge$_2$Te$_6$ contacts in the range of tens of nm, we consider them to be close to the bulk system~\cite{Rahman2021, Ko2022, Zhang2022}.
The CB minimum for Cr$_2$Ge$_2$Te$_6$ is located from 3.84 to 3.93~eV below the vacuum level~\cite{Kang2013, Gong2013, Rahman2021} and is slightly above the conduction band of \WSe ML~\cite{Rahman2021}. The reported calculations predict that the Cr$_2$Ge$_2$Te$_6$ band gap is around 200~meV~\cite{Ji2013, Yimei2018, Lin2017, Zhang2016, Li2018, Suzuki2019}, which positions the VB onset approximately 4.1 eV below the vacuum level. In our following analysis, we neglect that the presence of donor and acceptor states may lead to the occurrence of n- and p-doping in the Cr$_2$Ge$_2$Te$_6$ ~\cite{Ji2013, Tang2017, Tang2017, Hao2018, HATAYAMA2020, Rahman2021}. 

Under the resulting band alignment, the offset of the VB is about three times larger than the offset of the CB. The type II band alignment of the Cr$_2$Ge$_2$Te$_6$/hBN/\WSe/hBN/Cr$_2$Ge$_2$Te$_6$ heterostructure, with a few-layer hBN spacers, is pictorially presented in Fig.~\ref{fig_4}(b), and the resulting tunneling pathways for charge carriers upon application of a voltage are presented in Figs.~\ref{fig_4}(c,d). The application of voltage induces a tunneling current starting from almost 0~V, but an EL signal is observed at higher voltages of 7~V and -19~V. Due to the band alignment in our device, we propose that free electrons may tunnel immediately from the smallest voltages, as presented in Fig.~\ref{fig_4}(c). It is important to mention that the additional thermal energy of electrons at room temperature substantially facilitates their tunneling, likely due to the thermal excitation of impurity levels, as the corresponding tunneling current measured at 150~K is almost three orders of magnitude smaller compared to the one at 300~K (see the SI for the \textit{IV} comparison at different temperatures). At bias voltages on the order of 7~V and -19~V, holes start to tunnel, enabling the creation of electron and hole pairs, which leads to the radiative recombination, as schematically illustrated in Fig.~\ref{fig_4}(d).
For vdW LED devices with \WSe ML as an active medium and graphene layers as electrodes, voltages required to measure EL can be as small as the optical and electronic band gaps of the \WSe ML~\cite{Withers2015, Withers2015WSe2, Binder2017}. In our case, the much larger voltages needed to be applied for the EL can be described in the higher contact resistance~\cite{Ji2013, Zhang2016, WALCZYK2025}, in conjunction with different rates of Fermi level motion arising due to quantitatively different DOS between Cr$_2$Ge$_2$Te$_6$ and graphene. While the electronic structure of graphene is characterized by a linear dispersion relation near the Dirac points, resulting in the linear energy dependence of its DOS, the extrema of the CB and VB in the vicinity of the Cr$_2$Ge$_2$Te$_6$ band gap can be described by the parabolic dispersion (see the SI for details). The parabolic dispersion of electronic bands causes the square root dependence of DOS with a steep evolution at small energies, but also depends on the effective masses of electrons and holes, respectively, in CB and VB. Consequently, much larger voltages are required to observe the \WSe EL from the devices with the Cr$_2$Ge$_2$Te$_6$ electrical contacts.

% {Conclusions} %%%%%%%%%%%%%%%%%%%%%%%%%%%%%%%%%%%%%%%%%%%%%%%%%%
\section{Conclusions \label{sec:Conclusions}}
In summary, EL from a light-emitting tunneling structure based on a \WSe ML as an active emission material, with electrical contacts made of semiconducting Cr$_2$Ge$_2$Te$_6$, was investigated at room temperature (300~K). The EL intensity exhibited a non-monotonic evolution with applied voltage, while the tunneling current-voltage curve showed a gradual increase of the tunneling current.  This behavior can be explained by the formation of resonant and antiresonant tunneling conditions, which govern the competition of radiative and nonradiative tunneling pathways. This was evidenced by the integrated intensity of the EL signal, which qualitatively resembled the DOS of Cr$_2$Ge$_2$Te$_6$ calculated at the level of density functional theory. We found that the EL technique can be used to probe the DOS of a material acting as an electrical contact. These novel mechanisms of tunneling processes in vdW LEDs provide functionalities in the domain of electrically modulated optical devices operating at ambient conditions.

%{Methods} %%%%%%%%%%%%%%%%%%%%%%%%%%%%%%%%%%%%%%%%%%555
\section{Methods\label{sec:Methods}}

\subsection{Sample}
The examined LED was fabricated using mechanical exfoliation and dry transfer methods and placed on a silicon substrate covered by a thin SiO$_2$ layer. 
The heterostructure was composed of an ML of \WSe encapsulated between thin hBN barriers with top and bottom CGT electrodes, see Fig.~\ref{sheme}. 
As extracted by atomic force microscopy (AFM), the CGT contacts have different thicknesses of about 20 nm and 75 nm for the bottom and top contacts, respectively. 
The whole structure was additionally encapsulated between the hBN layers to protect against ambient conditions. 
After transfer, the gold contacts to the two CGT electrodes were attached with the aid of the ion-beam lithography technique.  

\subsection{Crystal growth}
To synthesize single CGT crystals, we employed a self-flux method, leveraging a germanium-tellurium rich flux, based on established procedures~\cite{Zeisner2019}.
All precursor handling occurred within an M-Braun glovebox, maintained in a rigorously controlled argon atmosphere (H$_2$O and O$_2$ levels below 0.1 ppm).
A mixture of chromium granules (4N purity, Thermo Fisher), germanium (4N purity, ChemPur), and tellurium lumps (5N purity, Alfa Aesar) was combined in a 10:13:77 molar ratio (Cr:Ge:Te) and loaded into an alumina crucible.
This crucible was then sealed within a 19 mm diameter fused-silica tube under a partial argon atmosphere (approximately 400 mbar).
The crystal growth process involved a carefully controlled thermal profile: the ampoule was heated to 1000$^{\circ}$C over 6 hours, kept at this temperature for 24 hours to ensure thorough mixing and dissolution, and subsequently cooled to 450$^{\circ}$C at a rate of 2$^{\circ}$C per hour.
At 450$^{\circ}$C, the excess Ge-Te flux was decanted by centrifugation, revealing plate-like millimeter-sized CGT crystals.
The crystal structure was determined using single-crystal X-ray diffraction, and the results, which align well with previously reported data~\cite{Carteaux1995, Yang2016}, are discussed in a different publication~\cite{Abadia-Huguet2025}.

%\textcolor{red}{What was the source of hBN and WSe2 crystals?}

\subsection{PL and EL measurements}
The PL experiments were performed using $\lambda = 561$ nm (2.21 eV) excitation from the continuous-wave (CW) laser diode. 
The studied sample was placed on a cold finger in a continuous-flow cryostat mounted on x–y motorized positioners. 
The excitation light was focused by means of a 50x long-working-distance objective with a 0.55 numerical aperture that produced a spot of about 1 $\mu$m diameter. 
The PL and EL signal was collected via the same microscope objectives, sent through a 0.75 m monochromator, and then detected using a liquid nitrogen-cooled charge-coupled device camera. 
Current-voltage characteristics were measured with a Keithley 2450 source meter.
The voltage was applied to the top electrode, while the bottom contact was grounded.

\subsection{Theoretical calculations}

DFT calculations were performed in Vienna ab-initio simulation package (VASP) version 6.4.2~\cite{KRESSE199615, PhysRevB.54.11169}. 
The projector augment wave (PAW) potentials Cr$_{pv}$, Ge$_{d}$ and Te with Pedew-Burke-Ernzerhof (PBE) parametrization of the general gradients approximation (GGA) to the exchange-correlation functional were used \cite{ PhysRevB.59.1758}. 
The unit cell parameters and atomic positions were optimized with criteria of 10$^{-5}$ ~eV/Å for forces and 0.1 kbar for stresses, including the D3 correction to van der Waals interactions \cite{10.1063/1.3382344}. 
An energy cut-off of 400 eV and a 6$\times$6$\times$6 G-centered k-points grid were used. 
The GGA+U approximation with effective Hubbard parameter U=1~eV was employed to properly model the ferromagnetic order of the magnetic moments of Cr atoms~\cite{ Kang2019}. 
DOS was calculated on a 12$\times$12$\times$12 k-points grid.

%%%%%%%%%%%%%%%%%% {Results} %%%%%%%%%%%%%%%%%

\section{Acknowledgment}
The work was supported by the National Science Centre, Poland (Grant No. 2022/46/E/ST3/00166, 2023/48/C/ST3/00309). 
T.W. gratefully acknowledges Poland's high-performance Infrastructure PLGrid ACC Cyfronet AGH for providing computer facilities and support within computational grant no. PLG/2025/018073. 
This project was supported by the Ministry of Education (Singapore) through the Research Centre of Excellence program (grant EDUN C-33-18-279-V12, I-FIM). This research is supported by the Ministry of Education, Singapore, under its Academic Research Fund Tier 2 (MOE-T2EP50122-0012). This material is based upon work supported by the Air Force Office of Scientific Research and the Office of Naval Research Global under award number FA8655-21–1-7026. 
K.W. and T.T. acknowledge support from the JSPS KAKENHI (Grant Numbers 21H05233 and 23H02052) and the World Premier International Research Center Initiative (WPI), MEXT, Japan.

\bibliographystyle{apsrev4-2}
\bibliography{biblio}

\newpage
\onecolumngrid

\include{si_arxiv}

\end{document}

%% file: si_arxiv.tex
\renewcommand{\thefigure}{S\arabic{figure}}
\renewcommand{\thesection}{S\arabic{section}}

\begin{center}
	%%%%%%%%% ABSTRACT TITLE
	{\large{ {\bf Supplementary Information \\for\\ Electrically modulated light-emitting diodes driven by resonant and antiresonant tunneling between Cr$_2$Ge$_2$Te$_6$ electrodes}}}
	%%%%%%%%% ABSTRACT AUTHORS
	\vskip0.5\baselineskip{Natalia Zawadzka,{$^{1}$} Kristina Vaklinova,{$^{2}$} Tomasz Woźniak,{$^{1}$} Mihai I. Sturza,{$^{3}$} Holger Kohlmann,{$^{3}$} Kenji Watanabe,{$^{4}$} Takashi Taniguchi,{$^{5}$} Adam Babiński,{$^{1}$} Maciej Koperski,{$^{2,6}$} and Maciej R. Molas{$^{1}$}}
	
	%%%%%%%%% AFFILIATION
	\vskip0.5\baselineskip{\em$^{1}$ University of Warsaw, Faculty of Physics, 02-093 Warsaw, Poland\\
		\em$^{2}$ Institute for Functional Intelligent Materials, National University of Singapore, 117544, Singapore \\
		\em$^{3}$ Institute of Inorganic Chemistry, Faculty of Chemistry and Mineralogy, Leipzig University, 04103 Leipzig, Germany 
		\em$^{4}$ Research Center for Electronic and Optical Materials, National Institute for Materials Science, 1-1 Namiki, Tsukuba 305-0044, Japan \\
		\em$^{5}$ Research Center for Materials Nanoarchitectonics, National Institute for Materials Science, 1-1 Namiki, Tsukuba 305-0044, Japan \\
		\em$^{6}$ Department of Materials Science and Engineering, National University of Singapore, 117575, Singapore}
	
\end{center}

\section{Low-temperature photoluminescence}

\begin{figure}[!h]
    \centering
    \includegraphics[width=0.5\linewidth]{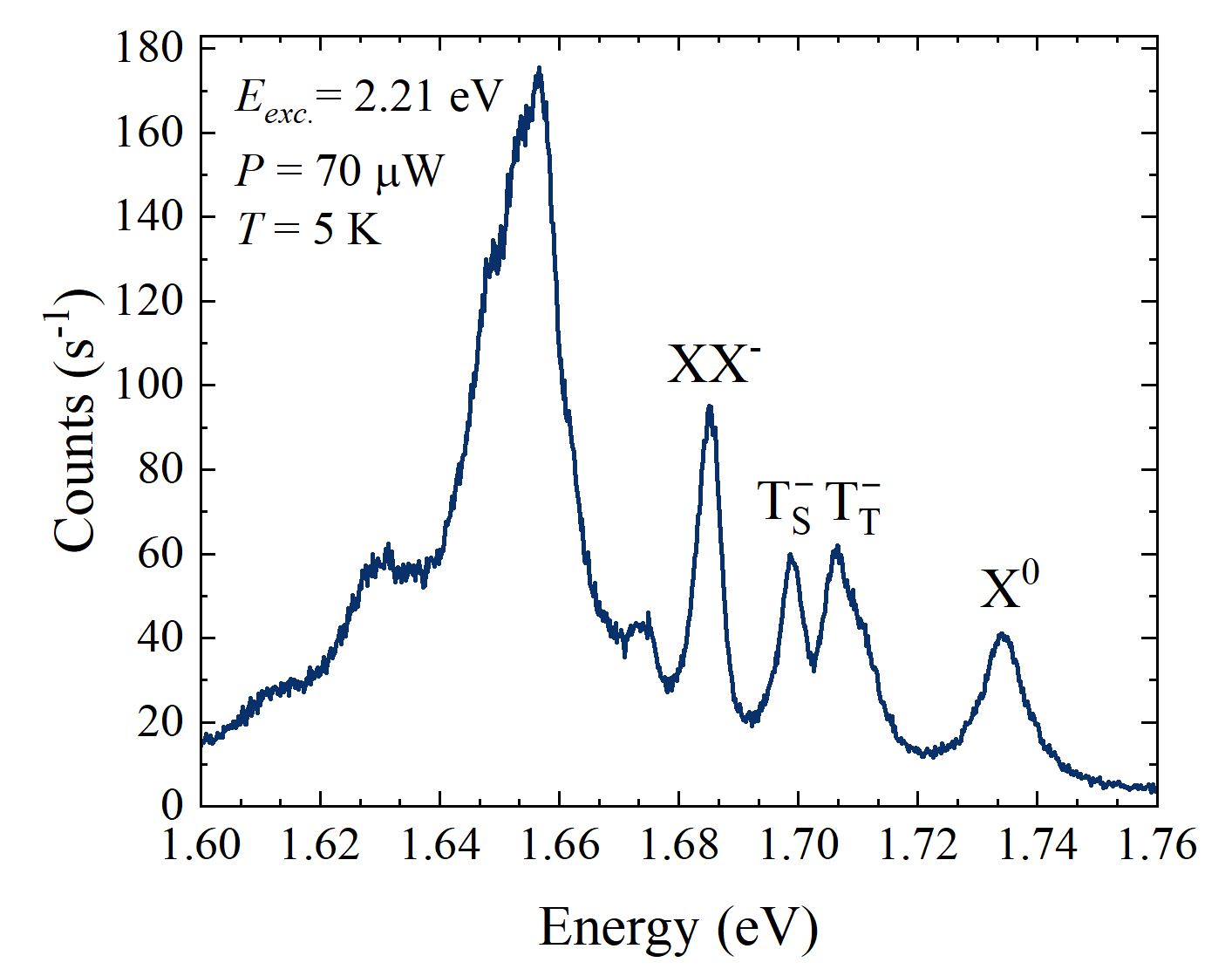}
    \caption
    {\label{SI_PL}
    PL spectrum of WSe$_2$ monolayer measured at 5~K under 2.21~eV and 70~$\mu$W laser excitation.  }
\end{figure}

A series of well-resolved emission lines seen in the low-temperature (5~K) PL spectra under 2.21~eV and 70~$\mu$W laser excitation confirms the high quality of the examined monolayer (ML); see Fig.~\ref{SI_PL}. 
Consequently, the assignment of the apparent peaks was performed according to the energy positions of excitonic complexes reported in the literature~~\cite{Courtade2017, Withers2015, Li2019, Zinkiewicz2022, WALCZYK2025}. 
The PL spectrum is composed of several emission lines ascribed to a neutral bright exciton (X$^0$), negatively charged biexciton (XX$^-$) and two bright negatively charged excitons (negative trions), $i.e.$ a spin–singlet (T$^-_S$) and a spin–triplet (T$^-_T$).
The emergence of the T$^-_S$ and T$^-_T$ lines in the measured PL spectrum confirms that the investigated WSe$_2$ ML is unintentionally n-doped, $i.e.$ there is a non-negligible concentration of free electrons. 

\clearpage
\section{Dependence of the charged and neutral exciton intensities on excitation power and bias voltage}

\begin{figure}[!h]
    \centering
    \includegraphics[width=0.8\linewidth]{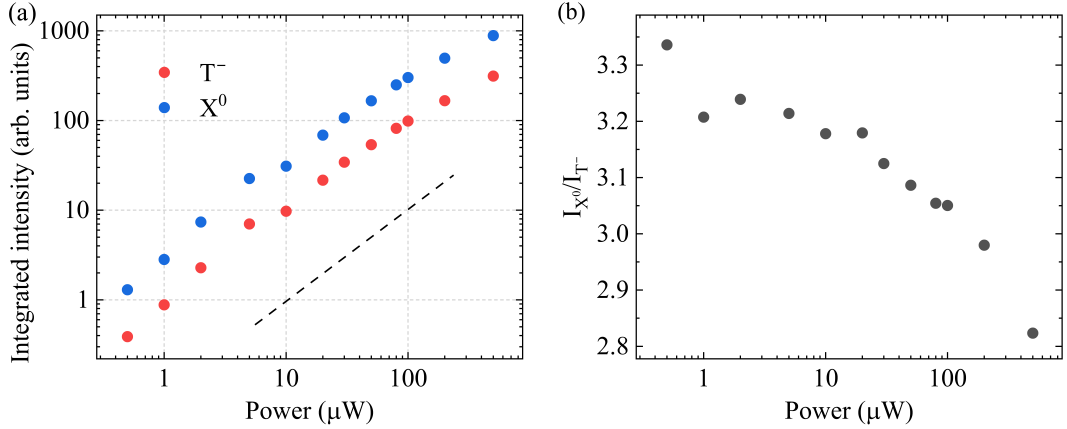}
    \caption
    {\label{SI_power} (a) Integrated intensities of the X$^0$ and T$^-$ lines as a function of excitation power.
    The dashed black line indicates linear behavior and serves as a guide to the eye.
    (b) Excitation-power dependence of the X$^0$/T$^-$ intensity ratio.}
\end{figure}

We measured the PL spectra of the WSe$_2$ MLs as a function of the excitation power at 300~K.
Figure~\ref{SI_power}(a) shows the evolution of the X$^0$ and T$^-$ emission intensities with excitation power. The sample was excited with an energy of 2.21 eV.
Both emission lines exhibit nearly linear dependencies, which are characteristic of free excitonic complexes.
The excitation-power dependence of the X$^0$/T$^-$ ratio shows a monotonic behavior (see Fig.~\ref{SI_power}(b)).
The ratio changes only about 10$\%$ even when the excitation power increases by four orders of magnitude.

\begin{figure}[!h]
    \centering
    \includegraphics[width=0.6\linewidth]{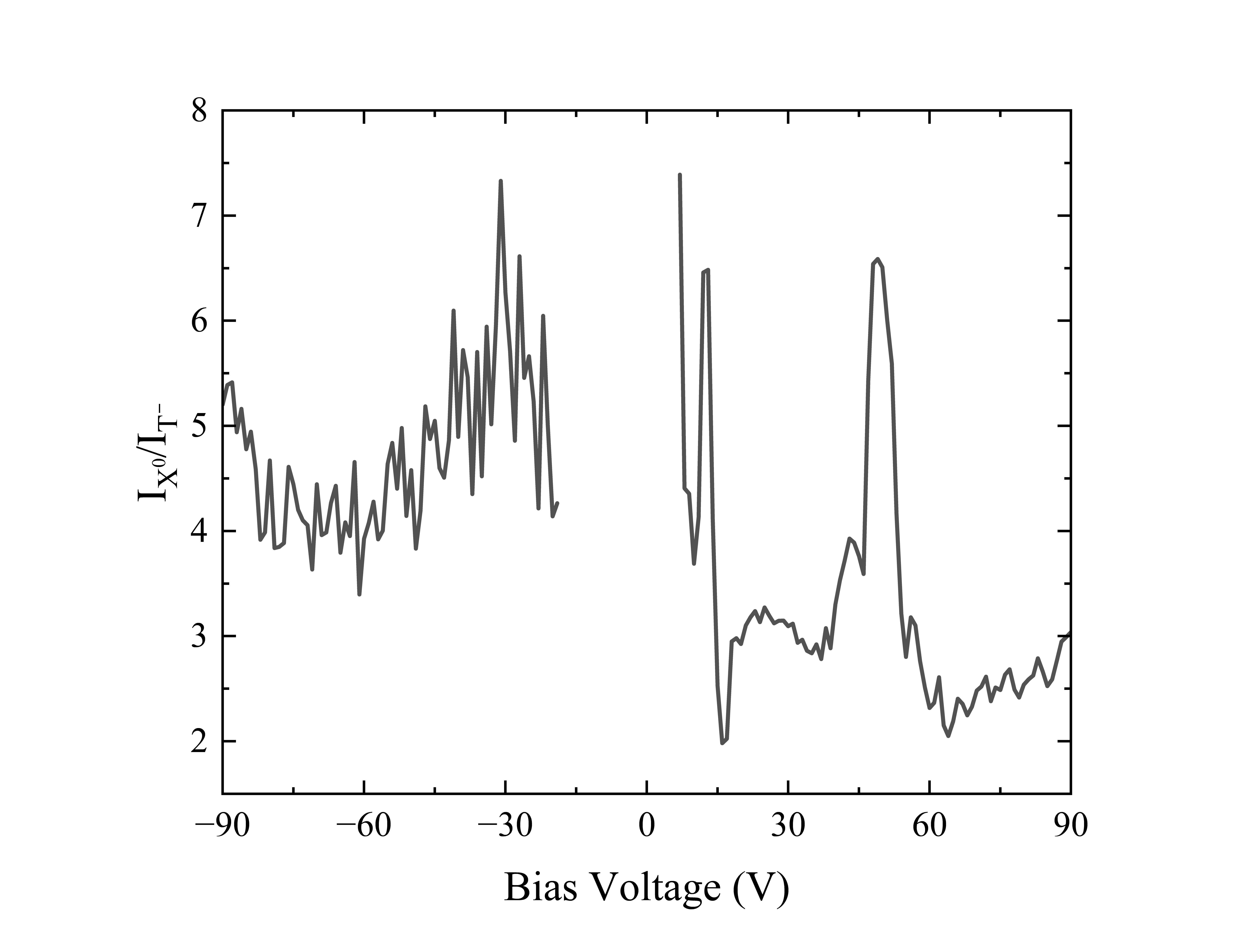}
    \caption
    {\label{SI_intensity_ratio_EL}
    Bias-voltage dependence of the X$^0$/T$^-$ intensity ratio.}
\end{figure}

The evolution of the X$^0$/T$^-$ intensity ratio with bias voltage at 300 K is shown in Fig.~\ref{SI_intensity_ratio_EL}.
The ratio exhibits a clear non-monotonic behavior, with pronounced changes particularly around 50 V, and varies between 2 and 7.4.

\clearpage
\section{Current-voltage curves}

\begin{figure}[!h]
    \centering
    \includegraphics[width=0.5\linewidth]{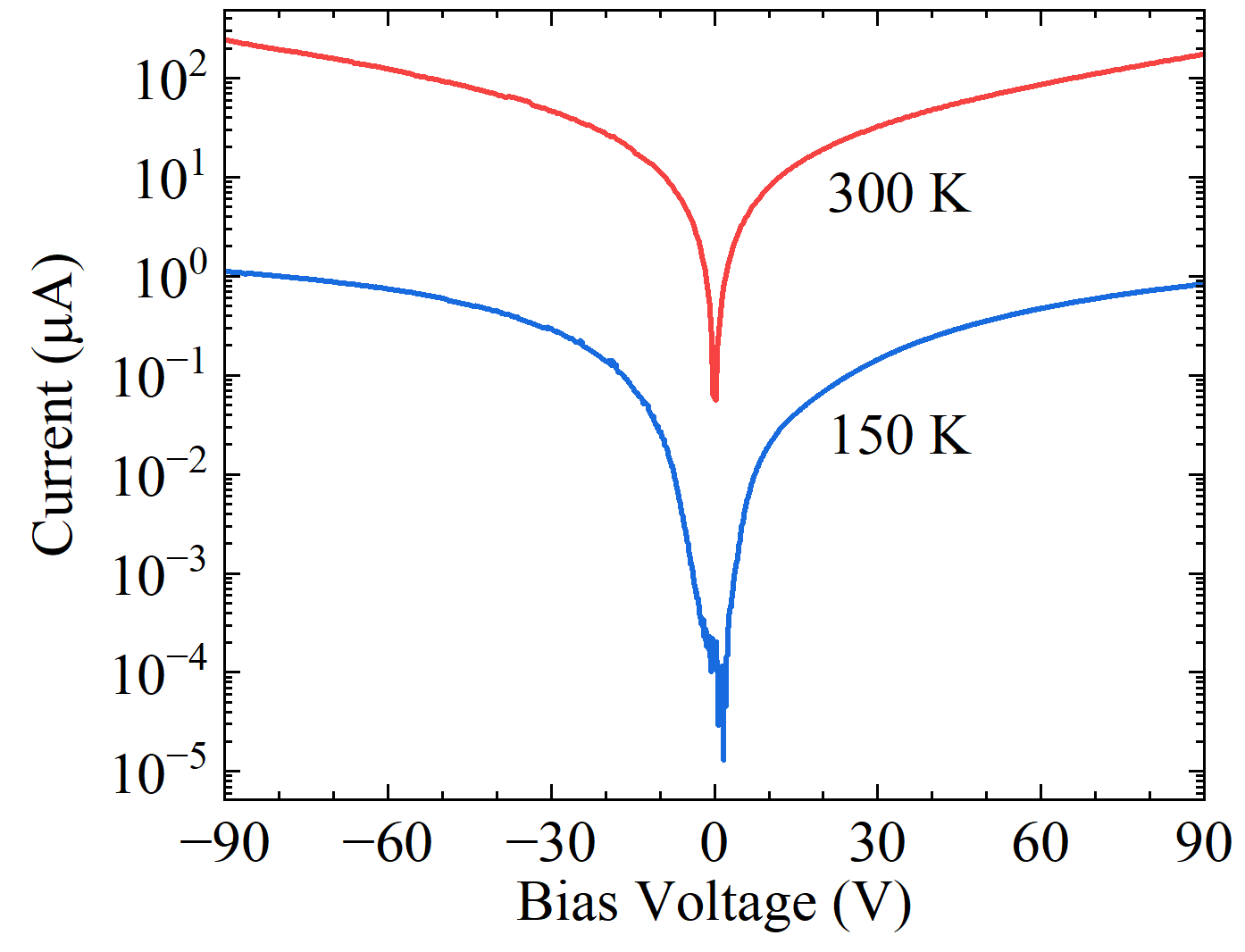}
    \caption
    {\label{SI_IV}
     The tunneling current–voltage curve measured on the studied LED device at 150~K and 300~K, presented on a logarithmic scale.}
\end{figure}

As mentioned in the main text, the optoelectronic properties of the LED studied are examined at room temperature ($T$=300~K).
The reason for this is due to the significant increase in the resistivity of CGT from about 0.46~$\Omega$/cm$^3$ at 300~K to 24~k$\Omega$/cm$^3$ at 50~K~\cite{Ji2013, Zhang2016}.
At lower temperatures, the resistivity is not reported because the sample resistivity is too high to measure~\cite{Ji2013, Zhang2016}.
This indicates that CGT is a good insulator below its ferromagnetic transition temperature at 60~K~\cite{Ji2013, Zhang2016}. 
However, the tunneling current-voltage (IV) curves at 150~K and at 300~K, presented in Fig.~~\ref{SI_IV}, show a typical shape for LED devices, the EL signal was only detected at room temperature.
At 300~K, the IV curve demonstrates a mono-exponential increase starting almost from 0~V, the corresponding EL signal is apparent at much higher voltages of about tens of volts. 
In contrast, at low temperature (150 K), the current starts to increase mono-exponentially when the applied voltage is about 4 V.
However, the EL signal was not observed at 150~K even at the highest applied voltages.
Probably, it is due to the fact that the corresponding tunneling current measured at 150~K is almost three orders of magnitude smaller compared to the one at 300~K. 
These results suggest that additional thermal energy of electrons at room temperature substantially facilitate their tunneling.

\clearpage
\section{Density of states of CGT and graphene}

\begin{figure}[!h]
    \centering
    \includegraphics[width=1\linewidth]{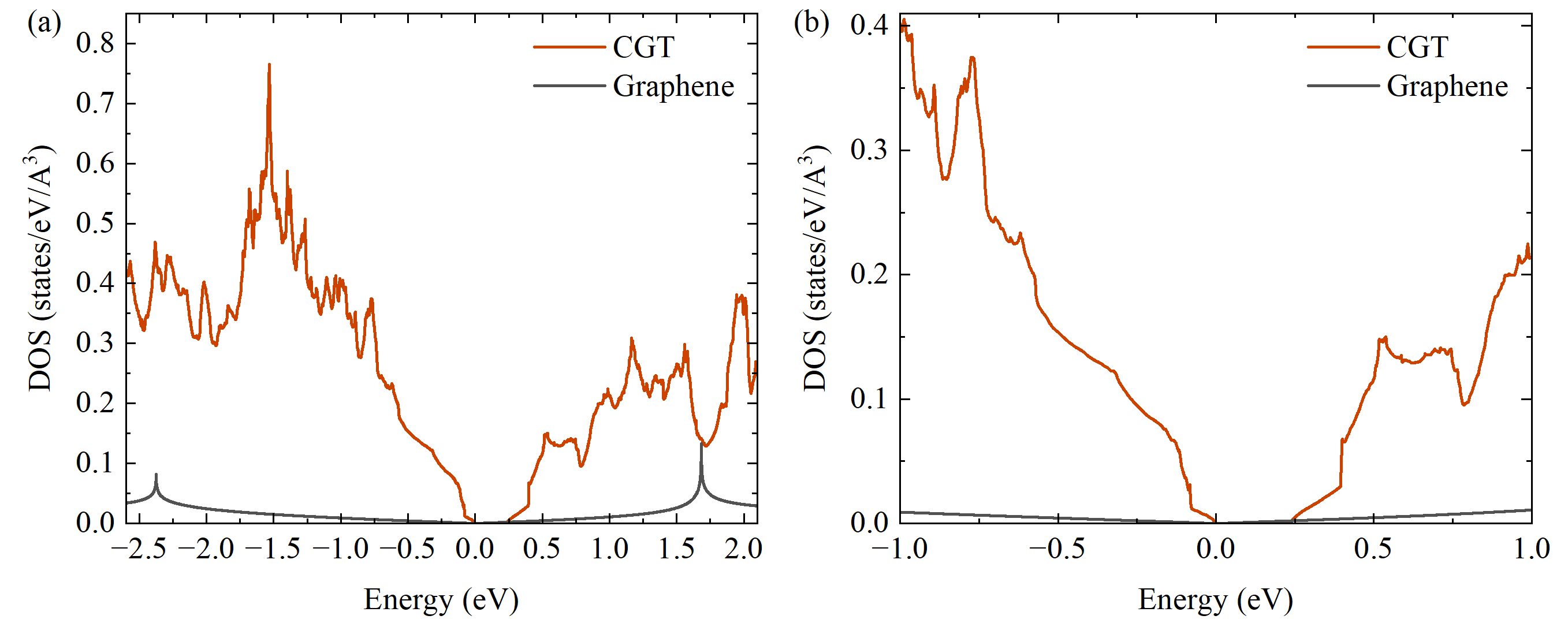}
    \caption
    {\label{SI_DOS}
    The calculated density of states of  (orange curve) CGT and (gray curve) graphene. DOSs are presented on different energy scales on two panels (a) and (b) to correspondingly emphasize the total energy range and the one in the vicinity of band gap/Dirac point.   }
\end{figure}

The comparison of the calculated DOSs of CGT and graphene is shown in Fig.~\ref{SI_DOS}.
As can be seen, the shape of the DOS of these two materials is different. 
The DOS of CGT is much larger as compared to the graphene one, see panel (a) of the Figure.
Moreover, the graphene DOS increases quadratically with energy, while the energy evolution of the CGT DOS is non-monotonic and complex, revealing its band structure (see next Section).
In the vicinity of the band gap and Dirac point of both materials, the DOS of CGT increases much faster with energy than that of graphene, which may have an impact when the CGT acts as an electrical contact. 
For example at 1~eV, the DOS of CGT is about 20~times bigger than DOS of graphene.

\clearpage
\section{Calculated electronic structure of CGT}

\begin{figure}[!h]
    \centering
    \includegraphics[width=0.7\linewidth]{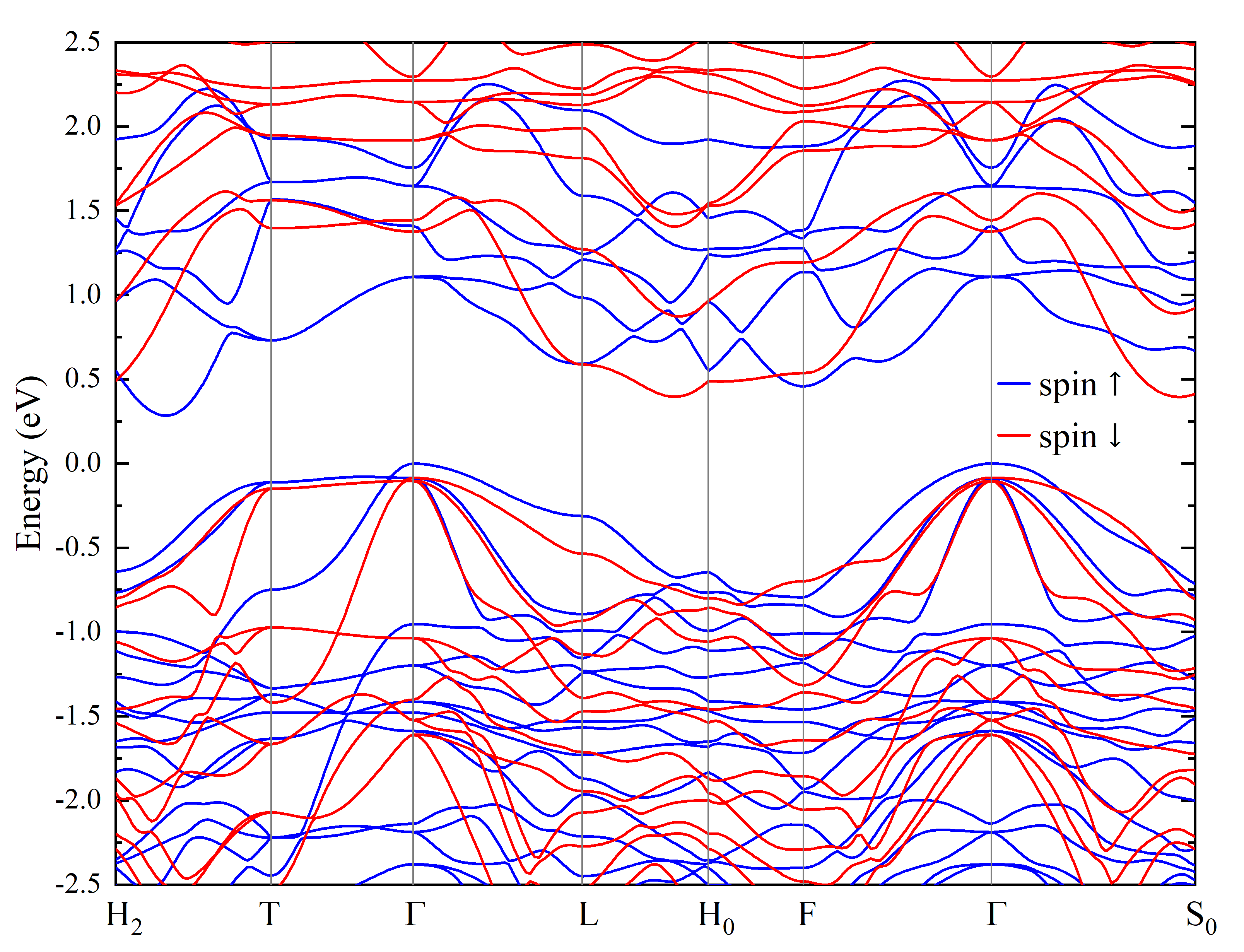}
    \caption
    {\label{SI_band}
    The calculated band structure of the CGT crystal in the ferromagnetic phase. The blue and red curves correspond to the spin up and down orientations, respectively.}
\end{figure}

The calculated electronic structure of the CGT crystal in the ferromagnetic phase is presented in Fig.~\ref{SI_band}.
The band structure of CGT is very complex with a tens of different subbands in both the conduction and valence bands, which may contribute to the tunneling process in the investigated LED device.

